# Decoherence of Dynamically Manipulated Qubits

**Vladimir Privman** and **Dmitry Solenov**

Center for Quantum Device Technology, Department of Physics,
Clarkson University, Potsdam, New York 13699-5721, USA

*Abstract:* We summarize our results on decoherence for short- to intermediate-time dynamics of an externally controlled two-state quantum system — a qubit — interacting with thermal bosonic environment. The developed approximation schemes are illustrated for an adiabatic model with time-dependent gate control, and for a model with rotating-wave gate function.

*Keywords:* decoherence, qubit, quantum computing.

## I. INTRODUCTION

Quantum information processing has been developing rapidly since the first algorithmic ideas that have clearly shown the power of quantum coherent computation [1,2]. Many physical realizations and conceptual models have been introduced since then [3-12]. For most of them, except perhaps for some resent proposals based on quantum walks [13,14], keeping the level of decoherence as low as possible helps achieve scalability [15-19]. Common quantum computation designs involve a sequence of "gate functions" corresponding to unitary operations performed on single qubits or on pairs of qubits, as well as measurements related to error correction [20]. Each gate function is carried out by applying external fields that control the dynamics of an individual qubit or modify an interaction between several qubits to entangle them [21].

In this work we focus on the single-qubit gate functions. For the unitary (coherent) dynamics without decoherence, the time-shape of the gate function is irrelevant as long as the final state has the desired form. It is however not the case when one accounts for quantum noise. The decohering effect of the environment can potentially be amplified as well as suppressed depending on the time-dependence of the applied gate potential. In particular, it was found that decoherence can be suppressed for qubit evolution subject to an appropriate sequence of rectangular pulses [22]. At the same time, it has been recognized that in many cases application of external control pulses can in itself lead to noise-like effects, such as destroying the qubit-qubit entanglement or "initializing" the qubits. Here we investigate the level of decoherence for single-qubit dynamics with nontrivially shaped gate functions and quantum noise induced by coupling to a thermal bath of bosonic modes [23].

The Hamiltonian of the gated qubit exposed to an environment (bath) of bosonic modes is

$$H = H_S(t) + H_B + H_{SB} . \quad (1)$$

Here $H_S(t)$ is the Hamiltonian of the qubit, including time-dependent gate control. The Hamiltonian of the environment is given by $H_B$, and $H_{SB}$ represents the interaction between the qubit and bath. The environment is often modeled as [24-25]

$$H_B = \sum_k \omega_k a_k^\dagger a_k , \quad (2)$$

where we set $\hbar = 1$. It is common [25] to assume the system-bath coupling in the form

$$H_{SB} = S \sum_k X_k , \quad (3)$$

where $S$ is a Hermitian operator in the system (qubit) space, and $\text{Tr } S = 0$, coupled to the bath operators $X_k$. We assume [25] linear coupling,

$$X_k = g_k a_k^\dagger + g_k^* a_k . \quad (4)$$

Initially, at time $t = 0$, the qubit is taken unentangled with the bath, which is considered a reasonable assumption when one is interested in the dynamics of decoherence [23,25]. The initial density matrix for the bath modes is fully thermalized: The overall density matrix at $t = 0$ is

$$\rho(0) = \rho_S(0) \frac{e^{-H_B/kT}}{\text{Tr}_B \left( e^{-H_B/kT} \right)} . \quad (5)$$

Since one of the criteria for designing a quantum computer is to minimize the noise effects, it is often expected that the interaction term is small with respect to other energy scales of the problem [18,19,26]. As a result, many approaches to evaluating decoherence utilize various perturbative approximations in powers of $|g_k|$. On the other hand, approximation techniques using the time itself as a small parameter once the fast dynamics has been factored out, have also been developed [27] and found many applications [28-31].

Here we report results within both types of approximation. In fact, the use of two different approximation schemes allows us to gauge the error of each of them and compare the physical phenomena that they capture. In Section II, we survey the derivation of the approximations and test them for a solvable model. Our main results are reviewed in Section III, where we consider the case of a qubit subject to a controlled, time-dependent gate function. We also analyze the nonlinear dependence of decoherence properties on the strength of the applied gate function. The perturbative approximation is shown to capture suppression of decoherence for a strong rotating-wave gate control.

## II. SHORT-TIME AND PERTURBATIVE EXPANSIONS

Let us review the two approximation techniques [23]. We begin with the assumption that the "ideal" evolution operator (without any noise), $U_S(t)$, of the coherent system, with the Hamiltonian $H_S(t)$, is known. The overall density matrix for $t > 0$ can be obtained in terms of the evolution operator, $U_T(t)$, defined by the Hamiltonian (1).

Within the short-time approximation [23,27], we approximately factor the evolution operator $U_T(t)$ into a product that allows us to compute the trace over the bath modes, preserving unitarity,

$$U_T(t) = W_S(t) e^{-i(H_B + H_{SB})t} W_S(t). \quad (6)$$

where the unitary operator $W_S(t) = \sqrt{U_S(t)}$ was introduced. Square root of a unitary operator is defined up to phases modulo $\pi$, in the diagonal representation. This ambiguity, however, does not affect the results. One can demonstrate [23] that approximation (6) is correct up to second order in time, inclusive. In the basis defined by the interaction, $S|\lambda\rangle = \lambda|\lambda\rangle$, the reduced density matrix can be found as

$$\rho_S(t) = \sum_{\lambda\lambda'} e^{D_{\lambda\lambda'}(t)} [W_S(t)|\lambda\rangle \\ \times \langle\lambda|W_S(t)\rho_S(0)W_S^\dagger(t)|\lambda'\rangle\langle\lambda'|W_S^\dagger(t)], \quad (7)$$

with the decoherence functions in the form

$$D_{\lambda\lambda'}(t) = -(\lambda - \lambda')^2 \sum_k \frac{2|g_k|^2}{\omega_k} \sin^2(\omega_k t/2) \coth(\omega_k/2kT) \\ - i(\lambda^2 - \lambda'^2) \sum_k \frac{|g_k|^2}{\omega_k} (\sin\omega_k t - \omega_k t). \quad (8)$$

Decoherence functions are usually evaluated in the continuum limit of infinitely many bath modes, by converting the summation over the modes to summation, and ultimately integration, over the mode frequencies. For our model calculations we will assume [25] the density, $\Upsilon(\omega)$, of the modes times the absolute square of the coupling constant, $|g(\omega)|^2$, to be proportional to a power-law function with an exponential cut-off, i.e., $\Upsilon(\omega)|g(\omega)|^2 = J\omega^n e^{-\omega/\omega_c}$.

The second approximation considered here is based on a perturbative expansion in terms of the qubit-bath coupling strength. It utilizes the Magnus expansion technique, [32], and is formulated to preserve the unitarity of the overall dynamics. We write the overall evolution operator in the interaction representation as an exponential of the Magnus series,

$$U(t) = \exp\left[-i\sum_{m=1}^\infty \Omega_m(t)\right]. \quad (9)$$

**Table 1**. Measure of decoherence calculated for the adiabatic model within the two approximation schemes considered. Ohmic ($n = 1$) and Super-Ohmic ($n = 2, 3$) cases are shown, for the parameter values $J = 10^{-6}$, $\omega_c = 30$, $T = 0$, $t = 1$.

|  | short-time approximation | perturbative approximation | the deviation |
|---|---|---|---|
| $n = 1$ | $3.48431 \times 10^{-6}$ | $3.48429 \times 10^{-6}$ | $2.41771 \times 10^{-11}$ |
| $n = 2$ | $3.06940 \times 10^{-5}$ | $3.06921 \times 10^{-5}$ | $1.87601 \times 10^{-9}$ |
| $n = 3$ | $9.22894 \times 10^{-4}$ | $9.21203 \times 10^{-4}$ | $1.69110 \times 10^{-6}$ |

where each element is a Hermitian operator. The $m^{th}$ element in the series is proportional to the $m^{th}$ power of the coupling strength. We truncate the expansion at the second order in $|g_k|$, keeping the first two terms,

$$\Omega_1(t) = \int_0^t dt' \, S(t') \sum_k X_k(t'), \quad (10)$$

$$\Omega_2(t) = -\frac{i}{2} \int_0^t dt' \int_0^{t'} dt'' [S(t') \sum_{k'} X_{k'}(t'), S(t'') \sum_{k''} X_{k''}(t'')]. \quad (11)$$

The operator (9) is then factorized into a product using the unitarity-preserving operator identities, see [23]. Finally, the reduced density matrix is found in the form

$$\rho_S(t) = \sum_{\mathbf{x},\mathbf{x}'} e^{D_{\mathbf{x},\mathbf{x}'}(t)} U_S(t) \mathfrak{M}_\mathbf{x} \rho_S(0) \mathfrak{M}_{\mathbf{x}'}^\dagger U_S^\dagger(t). \quad (12)$$

where $\mathfrak{M}_\mathbf{x} \equiv |x\rangle\langle x|y\rangle\langle y|z\rangle\langle z|$, and $|x\rangle, |y\rangle, |z\rangle$ are the eigenvectors of the corresponding Pauli matrixes. Here the decoherence functions are given by

$$D_{\mathbf{x},\mathbf{x}'}(t) = -\sum_k \frac{|g_k|^2}{2} \left[\{(\mathbf{x} - \mathbf{x}') \cdot \mathbf{f}_k\}^2 + \{(\mathbf{x} - \mathbf{x}') \cdot \tilde{\mathbf{f}}_k\}^2\right] \coth\frac{\omega_k}{2kT} \\ -i\sum_k |g_k|^2 \left[(\mathbf{x} - \mathbf{x}') \cdot (\mathbf{\gamma}_k - \mathbf{F}_k) + \{(\mathbf{x} - \mathbf{x}') \cdot \mathbf{f}_k\}\{(\mathbf{x} + \mathbf{x}') \cdot \tilde{\mathbf{f}}_k\}\right] \\ +i\sum_k |g_k|^2 \left[2(yx - y'x')f_k^y \tilde{f}_k^x + 2(zx - z'x')f_k^z \tilde{f}_k^x + 2(zy - z'y')f_k^z \tilde{f}_k^y\right]. \quad (13)$$

Where $\mathbf{F}_k \equiv \left(f_k^y f_k^z + \tilde{f}_k^y \tilde{f}_k^z, -f_k^x f_k^z - \tilde{f}_k^x \tilde{f}_k^z, f_k^x f_k^y + \tilde{f}_k^x \tilde{f}_k^y\right)$,

$$\mathbf{f}_k(t) = \int_0^t dt' \frac{Tr[S(t')\boldsymbol{\sigma}]}{2} \cos\omega_k(t' - t), \quad (14)$$

$$\mathbf{\gamma}_k(t) \equiv -\frac{i}{4} \int_0^t dt' \int_0^{t'} dt'' Tr(\boldsymbol{\sigma}[S(t'), S(t'')]) \cos\omega_k(t' - t'') \\ -\frac{1}{4} \int_0^t dt' \int_0^{t'} dt'' Tr(\boldsymbol{\sigma}\{S(t'), S(t'')\}) \sin\omega_k(t' - t''), \quad (15)$$

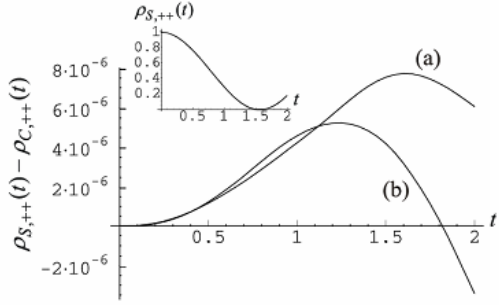

**Figure 1.** The inset shows the probability for the "ideal" system initially in the state $|+\rangle$ to remain (be measured) in the same state at later times. With the quantum noise, this probability deviates from the ideal. This deviation is shown calculated (a) within the short-time approach, and (b) whitin the perturbative approximation. The parameters are the same as in Table 1.

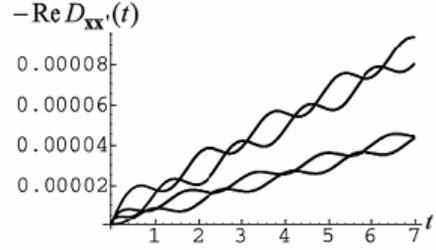

**Figure 2.** The real (damping) part of the decoherence functions obtained within the perturbative approach. The parameters are the same as in Figure 1. Development of two decoherence time scales is observed.

while the functions $\tilde{f}_k^j(t)$ are obtained from (14) by replacing the cosine with sine in the integral. This perturbative approximation can be easily tested for the exactly solvable adiabatic model [23], for which the first, short-time approximation coincides with the exact solution. In Table 1, we compare the measure of decoherence [33] calculated within both approaches for such an adiabatic model. The measure is calculated based on the deviation of the reduced density matrix from the ideal (coherent) density matrix,

$$\chi(t) = \rho_S(t) - \rho_C(t). \quad (16)$$

It is defined in terms of the eigenvalues $\varsigma_j$ of $\chi(t)$, i.e.,

$$\|\chi(t)\|_\lambda = \max_j |\varsigma_j(t)|. \quad (17)$$

We note that the quantity (17) is often maximized over all possible initial states to eliminate the qubit-frequency oscillatory behavior, in order to obtain an upper-bound estimate of decoherence. For our purposes it is sufficient to see the difference between the two approximations for a single, randomly chosen initial state.

### III. QUBIT SUBJECT TO A TIME-DEPENDENT GATE

Let us now consider a phase plus NOT gate combination [20,23,34,35]: We take a resonant rotating wave gate function as a simple example. The qubit part of the Hamiltonian then becomes

$$H_S(t) = a\sigma_z + c\left(\sigma_x \cos 2at + \sigma_y \sin 2at\right), \quad (18)$$

where $a$ is half the energy gap of the idling qubit, and $c$ is the amplitude of the gate function. The evolution operator for the Hamiltonian (16) can be derived analytically [23],

$$U_S(t) = e^{-iat\sigma_z} e^{-ict\sigma_x}. \quad (19)$$

One can easily verify that by applying the above gate function for $t = \pi/2a$, for example, with the amplitude $c = a$, the state $A|+\rangle + B|-\rangle$ is changed to $A|-\rangle - B|+\rangle$. Here $|\pm\rangle$ are the eigenstates of $\sigma_z$.

Let us first consider the effect of the noise on the oscillations between the states $|\pm\rangle$ due to (18). In Figure 1, we plot the deviation of the probability $\rho_{S,++}(t) \equiv \langle+|\rho_S(t)|+\rangle$ from the same coherent quantity. The approximations agree well for almost a full cycle of the gate. In fact, the short time approximation is consistent with the perturbative one up to the times of order $1/a$ or $1/c$, whichever is smaller, as expected from the range-of-validity analysis of the short-time approximation given in [23].

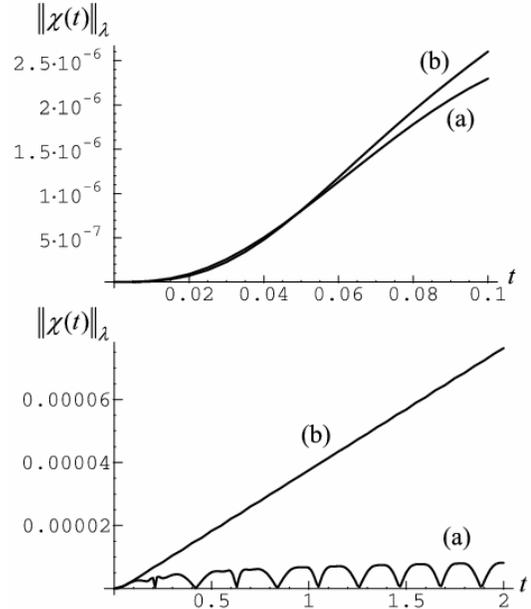

**Figure 3.** The measure of decoherence (17) for a large-amplitude gate function, $c = 15a$. The results for the short-time approach, (a), and perturbative technique, (b), are given. (The top panel shows these results for the initial times.) Other parameters are the same as in Figure 1. The decoherence effects are greater than for $c = a$, see Figure 1. Furthermore, noncummutativity effects set in earlier, as seen by the divergence of the curves for the two approximations.

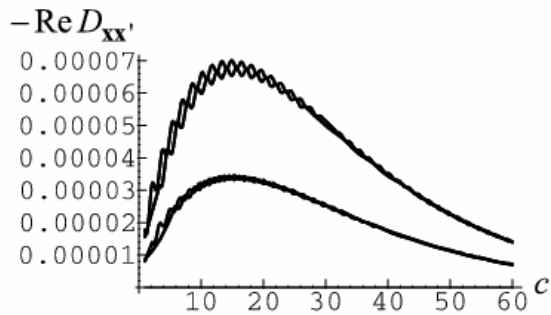

**Figure 4.** The decoherence functions of the perturbative approximation calculated for different values of the gate amplitude, $c$, at time $t = 1$. Other parameters are the same as in Figure 3. For large $c$, decoherence drops down almost to the level of the "weak" gate, see also Figures 1 and 2.

To understand the time scales of the decoherence processes let us analyze the decoherence functions (13) for the same system, see Figure 2. The real parts of the nonzero $D_{xx'}(t)$ separate into two bands growing on average linearly with time, each with superimposed oscillations. This defines two different time scales for the decoherence processes. Note that up to times for which the approximations agree, see Figure 1, there is no clear separation in the decoherence functions. At later times, however, the non-commutativity of the system Hamiltonian with the interaction becomes important, and the curve calculated with the short-time approach, which treats non-adiabatic terms as a "correction," departs from the one obtained within the perturbative expansion.

Our focus now will be on the development of the decoherence for different amplitudes of the applied gate function. In Figure 3, we plot the measure of decoherence (17) for the same set of parameters used in Figure 1 and the gate amplitude $c = 15a$. As seen from the figure, the decoherence here is much more significant than for the case $c = a$.

However, decoherence does not grow monotonically with $c$. The analysis of the decoherence functions for different gate amplitudes, see Figure 4, shows that at some point the influence of the gate on decoherence starts decreasing. Both bands of the nonzero decoherence functions increase with increasing gate amplitude, but only until the latter reaches approximately half of the bath cut-off frequency. At that point they begin decreasing, as do the imaginary parts of the decoherence functions, not shown in Figure 4, see [23]. Such a phenomenon can be explained as follows. As the system dynamics, driven by the external wave, becomes faster than the cutoff frequency of the bath modes, the resonant energy-exchange relaxation processes are suppressed. The pure-decoherence type noise still remains, and is captured correctly by the developed approximation schemes.

In summary, we surveyed novel approximation schemes for evaluating decoherence for time-dependent gate functions. The case of a qubit subject to a rotating wave gate function was studied in detail. Our findings indicate that the influence of quantum noise on the qubit dynamics is nontrivially affected by the internal structure of the gate used to manipulate the qubit.

The authors acknowledge support of this research by the NSF under grant DMR-0121146.